\documentclass[fleqn,usenatbib]{mnras}

\usepackage[english]{babel}
\usepackage[usenames]{color}

\usepackage{newtxtext,newtxmath}
\usepackage[T1]{fontenc}
\usepackage{graphicx}	% Including figure files
\usepackage{amsmath}	% Advanced maths commands

\title[Primary disks and their observational appearance]{Primary disks and their observational appearance in collapsing
magnetic rotating protostellar clouds}

\author[N. S. Kargaltseva, et al.]{
Natalya S. Kargaltseva$^{1,2}$\thanks{E-mail: kargaltsevans@mail.ru},
Sergey A. Khaibrakhmanov$^{1,2}$\thanks{E-mail: khaibrakhmanov@csu.ru},
Alexander E. Dudorov$^{2,1}$\thanks{deceased},
Andrey G. Zhilkin$^{3}$
\\
$^{1}$Ural Federal University, 51 Lenin st., Ekaterinburg 620000, Russia\\
$^{2}$Chelyabinsk State University, 129 Br. Kashirinykh st., Chelyabinsk 454001, Russia\\
$^{3}$Institute of Astronomy of the Russian Academy of Sciences, 48 Pyatnitskaya st., Moscow 119017, Russia
}

\date{Accepted 08.07.2021}
\pubyear{2021}

\begin{document}
\label{firstpage}
\pagerange{\pageref{firstpage}--\pageref{lastpage}}
\maketitle

\begin{abstract}
The collapse of the magnetic rotating protostellar cloud with mass of $10\,M_{\odot}$ is numerically studied.
The initial ratios of the thermal, magnetic, and rotational energies of the cloud to the modulus of its gravitational energy are 0.3, 0.2 and 0.01, respectively. The emphasis is on the evolution and properties of the
quasi-magnetostatic primary disk formed at the isothermal stage of the collapse. Simulations
show that the primary disk size and mass increase during evolution from $1500$~au to $7400$~au and from $0.3\,M_{\odot}$ to $5.2\,M_{\odot}$, respectively. Magnetic field is quasi-radial in the cloud envelope and quasi-uniform within the primary disk. A toroidal magnetic field is generated behind the front of the fast shock MHD wave propagating from the primary disk boundary and in the region of the outflow formed
near the first hydrostatic core. The hierarchical structure of collapsing protostellar clouds can be
revealed in observations in terms of the magnetic field geometry and the angular momentum distribution.
\end{abstract}


\section{Introduction}
\label{sect:intro}

The current star formation occurs in magnetic rotating cores of molecular clouds, i.e., protostellar
clouds (hereafter, PSCs). During the gravitational collapse of the PSC, a protostar is formed at its center,
which is observed as an infrared source~\citep[see][]{Zhang2020}.
The young protostar immersed in an extended envelope
is observed in the submillimeter range as a Class 0 young stellar object (hereafter, YSO). Characteristic
features of Class 0 YSOs are outflows~\citep{andre1993}. In Class 0 YSOs, flattened envelopes with radii of$200-10000$~au are observed, as well as small probably Keplerian disks with radius of $5-50$~au~\citep{ohashi97, wiseman2001}. In YSO envelopes, a large-scale magnetic field with hourglass geometry is detected; within disks, a pinch magnetic field and indications of a toroidal magnetic field are observed~\citep[see][]{lee2019}. 
The angular momentum distribution changes in going from the disk to the envelope~\citep{caselli02}.

The first simulations of the PSC collapse showed that the collapse is inhomogeneous with the formation of the first hydrostatic core at the cloud center~\citep{larson1969}. 
During the collapse, the magnetic rotating PSC takes a shape oblate along magnetic field lines and/or the rotation axis~\citep{scott1980}. The PSC magnetic flux evolution and properties of formed stars are substantially controlled by ionization, recombination, and diffusion MHD effects, in particular, ambipolar diffusion~\citep{DudSaz1987}.

The basic questions in the star formation theory are the problems of the angular momentum and catastrophic magnetic braking~\citep{scott1980, DudSaz1982}. The modern numerical simulations are mostly devoted to the accretion stage of the solar-mass PSC collapse~\citep[see][]{HennFromang2008, Zhao2020}.
To solve the angular momentum problem, it is important to comprehensively study
initial PSC collapse stages, when magnetic braking is the most efficient.

Previously, ~\cite{paper1} studied the isothermal collapse of magnetic PSCs with masses of $1$ and $10\,M_{\odot}$. Simulations showed that a hierarchical PSC structure is formed during isothermal collapse, which consists of a geometrically thick and optically thin envelope, with a geometrically and optically thin quiasi-magnetostatic primary disk (hereafter PD) inside.

The PD boundary is characterized by a sharp jump in velocity profiles, when almost free fall of the gas from
the cloud envelope to its center transforms to slow almost radial motion. At the PD boundary, a fast
shock magnetohydrodynamic (hereafter MHD) wave is formed, which moves to the cloud periphery.

In the present study, the approach of~\cite{paper1} is developed. The collapse of the magnetic rotating PSC with mass of $M_{\odot}$ is numerically simulated taking into account the formation of the first core. The PD evolution is studied, its mass, size, angular momentum, magnetic flux, and lifetime are
determined. Possible observational manifestations of PDs are discussed.

\section{Problem statement and numerical method}
\label{sect:problem}

A homogeneous spherically symmetric rotating PSC with mass of 10 $M_{\odot}$ and a temperature of 20 K, in a uniform magnetic field is considered. The initial cloud density is $4\cdot 10^4$~cm$^{-3}$, the initial cloud radius is $0.1$~pc. The main parameters determining the collapse dynamics are the ratios of the thermal $\varepsilon_{\rm t}$, magnetic $\varepsilon_{\rm m}$ and rotational $\varepsilon_{\rm w}$ energies to the modulus of its gravitational energy. In this paper, we consider the simulation
with $\varepsilon_{\rm t}= 0.3$, $\varepsilon_{\rm m}=0.2$ and $\varepsilon_{\rm w}=0.01$. 

The PSC collapse is studied using gravitational MHD equations. Numerical simulation is performed
using the two-dimensional MHD code `Enlil'~\citep{Dud1999, zhilkin09}. The PSC thermal evolution is simulated using the gas law with a density-dependent effective adiabatic index, $\gamma_{\rm eff}$~\citep{MasunInuts2000}. For the isothermal collapse $\gamma_{\rm eff}=1.001$ is taken. At the density $\rho\geq 10^{-13}$~g~cm$^{-3}$, when the first hydrostatic core is formed~\citep{larson1969}, $\gamma_{\rm eff}=5/3$. This approach allows us to model the PSC collapse taking into account the first core formation.

\section{Evolution of the primary disk during the protostellar cloud collapse}
\label{sect:fiduc_m}

The performed simulations confirm the conclusions by~\cite{paper1}. At the isothermal
collapse stage, the cloud gains a hierarchical structure: the envelope takes a shape oblate along magnetic
field lines and the rotation axis; a quasi-magnetostatic PD is formed within it. Let us consider the general
picture of the PD evolution with the emphasis on the angular momentum distribution.

\begin{figure*}
%\setcaptionmargin{5mm}
   \centering
  \includegraphics[width=0.79\textwidth]{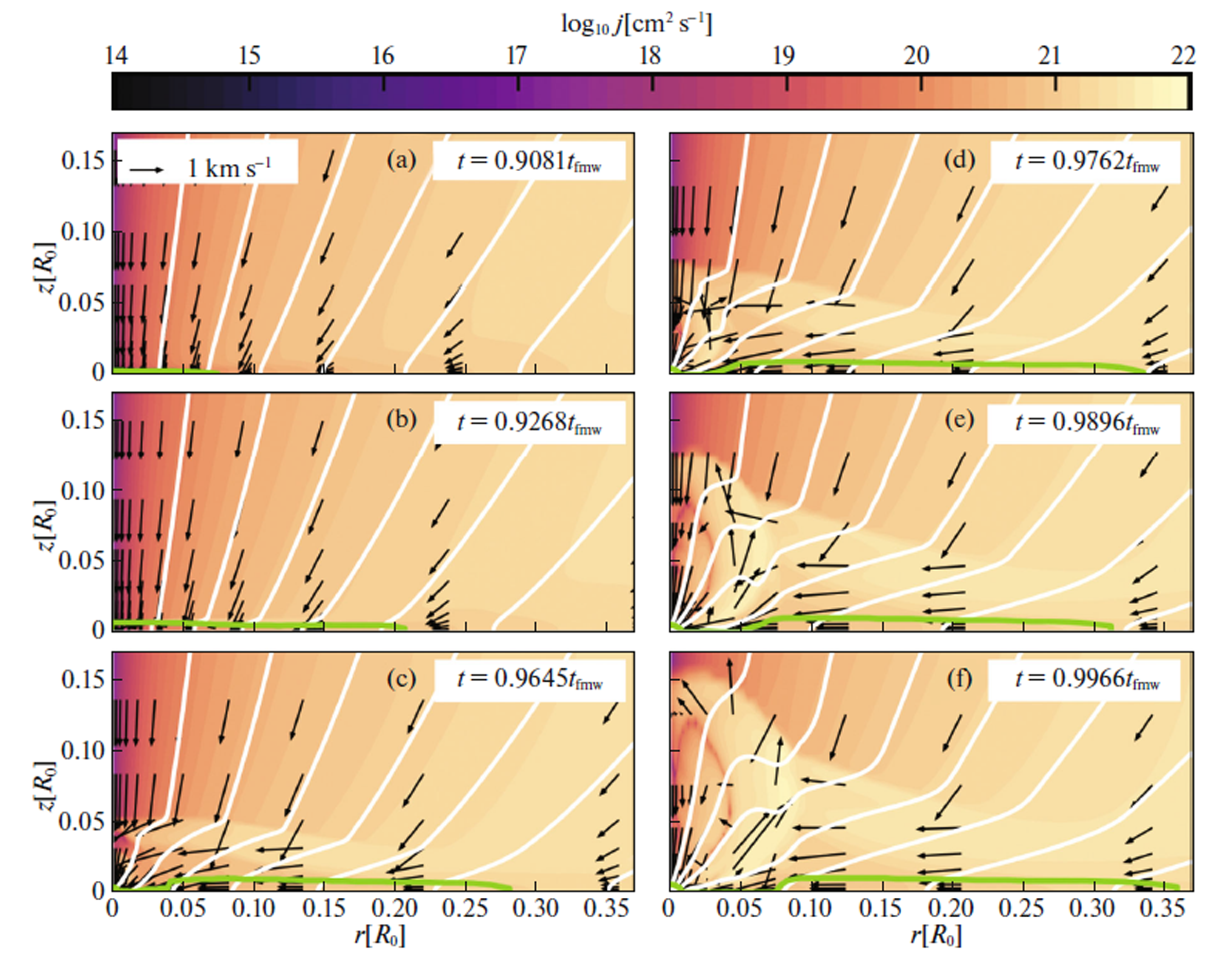}
   \caption{Distribution of the angular momentum (color filling), velocity field (arrows), and magnetic field (white lines) near
the primary disk at following time moments: a) $t=0.9081$ $t_{\rm fmw}$; b) $t=0.9268$ $t_{\rm fmw}$; c) $t=0.9645$ $t_{\rm fmw}$; d) $t=0.9762$ $t_{\rm fmw}$; e) $t=0.9896$ $t_{\rm fmw}$; f) $t=0.996$ $t_{\rm fmw}$. The green line is the primary disk boundary.} 
   %\captionstyle{normal}
   \label{fig:J_PD}
\end{figure*}

Figure~\ref{fig:J_PD} shows the quarter of the central part of the collapsing PSC at different time points. The time $t$ is measured in the units of the characteristic collapse time taking into account the effect of electromagnetic and centrifugal forces: $t_{{\rm fmw}} = t_{{\rm ff}}(1-\varepsilon_{\rm m}-\varepsilon_{\rm w})^{-1/2}$, where $t_{{\rm ff}}$ is the free fall time~\cite{DudSaz1982}. At the PD formation
time, $t=0.9081$ $t_{{\rm fmw}}$ (Fig.~\ref{fig:J_PD}a), its radius is $R_{\rm pd}\approx 0.07\,R_0\approx 1500$~au, and the ratio of its maximum half-thickness to the radius is $Z_{\rm pd}/R_{{\rm pd}}=0.039$. 
At the time point $t=0.9268$ $t_{{\rm fmw}}$ (Fig.~\ref{fig:J_PD}b), the PD radius is $\approx 0.22\,R_0\approx 4500$~au. At the PD boundary, a fast MHD shock wave is formed, which propagates into the
cloud envelope~\citep[see][]{paper1}.
The first core is formed at time $t=0.9645$ $t_{{\rm fmw}}$ (Fig.~\ref{fig:J_PD}c). Near the core,
in the region $r<0.04\,R_0\approx 800$~au, the PD thickness abruptly decreases, i.e., quasi-magnetostatic equilibrium is broken. Then, gas in this region begins to move from the cloud core to the periphery in parallel to the rotation axis, i.e., the outflow is formed (Fig.~\ref{fig:J_PD}d). The PD radius continues to increase and
by the time of $t=0.9966$~$t_{{\rm fmw}}$ (Fig.~\ref{fig:J_PD}f) reaches $R_{\rm pd}\approx 0.36\,R_0\approx 7400$~au. During evolution, the PD becomes
geometrically thinner $Z_{\rm pd}/R_{{\rm pd}}=0.028$, and the outflow region sizes increase.

At the PD formation time (Fig.~\ref{fig:J_PD}a) the magnetic field is quasi-radial, $B_r\sim B_z$,in the envelope and
quasi-uniform, $B_r\ll B_z$, within the PD. Further the magnetic field takes a toroidal geometry, $B_{\varphi}\sim(B_r, B_z)$, behind the front of the fast shock MHD wave and in the outflow region.

After the PD formation, the specific angular momentum is accumulated at its boundary (Fig.~\ref{fig:J_PD}b), and it is further transferred from the PD boundary to the PSC envelope by the fast shock MHD wave
(Fig.~\ref{fig:J_PD}c). The angular momentum of the first core is removed by the outflow (Figs~\ref{fig:J_PD}d--f).

An analysis of the simulation shows that the PD mass increases with time from $0.3\,M_{\odot}$ to $5.2\,M_{\odot}$. On the contrary, the PSC envelope mass decreases from the time of the PD formation. The total angular
momentum of the PSC decreases by 15~\% relative to the initial value~$J_0$.

\section{Conclusions and discussion}
\label{sect:end}

The present paper is the development of the study by~\cite{paper1}, in which the isothermal stage of the PSC collapse was modelled.
Two-dimensional numerical MHD simulation of the collapse of the magnetic rotating PSC with mass of 10~$M_{\odot}$ was performed taking into account the first hydrostatic core formation.

The simulations show that the hierarchical structure of  the PSC formed at the isothermal collapse stage~\cite{paper1}, is retained during further evolution. The first hydrostatic core is formed at the center of the quasi-magnetostatic PD. Near the core, $r<0.04\,R_0\approx 800$~au, the quasi-magnetostatic equilibrium is broken, and, afterwards,
outflow arises. The PD size and mass increase from 1500~au to 7400~au and from 0.3~$M_{\odot}$ to 5.2~$M_{\odot}$, respectively.
These values are close to characteristics of observed Class 0 YSO envelopes~\citep{ohashi97, wiseman2001}. Therefore, it can
be assumed that the observed large-scale oblate envelopes of Class 0 YSOs are PDs.

Our results confirm the conclusions by ~\cite{paper1} on the PD lifetime and
magnetic field geometry in the collapsing PSC. The PD is a long-living structure which continues to
evolve after the first core formation. The magnetic field geometry is different across the hierarchy. The magnetic field is quasi-radial within the envelope; it is toroidal behind the front of the fast MHD wave
emerging from the PD soon after its formation and in the outflow region; the
magnetic field is quasi-uniform within the PD. The PD plays an important role in the evolution of the specific angular momentum
in the cloud. The angular momentum is transferred by the fast shock MHD wave travelling
from the PD boundary to the cloud envelope, as well as by the outflow formed near the first core. The total
angular momentum of the cloud, decreased by 15~\% relative to the initial value by the time $t=0.9645$ $t_{{\rm fmw}}$ $=0.1936$ Myrs, when a typical hierarchical PSC structure with outflows.

Based on our results, it can be assumed that the hierarchical structure of collapsing PSCs can
be revealed in observations in terms of the distribution of the magnetic field geometry and the angular momentum.
To determine the magnetic field geometry at various hierarchy levels, polarization maps of Class 0 YSOs
should be constructed in the submillimeter range with high spatial resolution. 

The further study will be directed to the determination of the mass, size, total angular momentum, and magnetic flux of the PD, as well as of the first hydrostatic core during the collapse of magnetic rotating PSCs with various initial cloud parameters and taking into account magnetic field diffusion.

{\bf Acknowledgments}
This study was supported by the Russian Science Foundation, project no. 19-72-10012. The authors are grateful to reviewer E.O. Vasil’ev for helpful comments.

\bibliographystyle{mnras}
\bibliography{bib} 

\end{document}